\documentclass[prd,twocolumn,showpacs,floatfix,amsmath,amssymb,floatfix]{revtex4}
\usepackage{graphicx,color,dcolumn,booktabs,bm}
\usepackage{longtable,lscape}
\usepackage{txfonts}
\usepackage{overpic}
\usepackage{longtable, booktabs}
\usepackage{amssymb}
\usepackage{indentfirst,,multirow}
\usepackage{indentfirst}
\usepackage{feynmf}   
\usepackage{slashed}  
\usepackage{cases}
\usepackage{color}

\definecolor{mygray}{gray}{.8}
\definecolor{mypink}{rgb}{.99,.70,.95}
\definecolor{mycyan}{cmyk}{.3,0,0,0}
\definecolor{mygreen}{cmyk}{.4,0,0.5,0}
\definecolor{myyellow}{cmyk}{0,0,0.3,0}

\graphicspath{{Figures/}} %

\begin{document}

\title{An overview of $XYZ$ new particles}
\author{Xiang Liu$^{1,2}$}\email{xiangliu@lzu.edu.cn}
\affiliation{$^1$School of Physical Science and Technology, Lanzhou University,
Lanzhou 730000, China\\
$^2$Research Center for Hadron and CSR Physics,
Lanzhou University $\&$ Institute of Modern Physics of CAS,
Lanzhou 730000, China}

\begin{abstract}

In the past decade, more and more charmoniumlike and bottomoniumlike states have been reported in experiments, which have led us to extensive discussions on the underlying structure of these states. In this review paper, we briefly summarize the experimental and theoretical status of these observed states.

\end{abstract}
\pacs{***} \maketitle

\section{introduction}\label{sec1}

As the theory of describing the strong interaction, quantum chromodynamics (QCD) has made a remarkable success in interpreting hadron physics. In the QCD theory, high energy behaviors corresponding to short-distance interaction are
quite different from low energy behaviors that are determined by the color confinement. In the case of high energy processes, strong interaction is well depicted by the perturbation theory due to
the asymptotic freedom. However, for the low energy processes which are completely governed by nonperturbative
QCD effects, the situation becomes complicated and difficult since there is a lack of any reliable approach to deal with the QCD nonperturbative problem.
The lattice QCD theory is the one way to well treat nonpertubative pheonmena but it has just begun to explain many of these phenomena. Thus, it is an interesting and important research topic in hadron physics to search for a suitable way to quantitatively describe the color confinement and its results.

\begin{center}
\begin{figure}[htb]
\begin{tabular}{cccc}
\scalebox{0.6}{\includegraphics{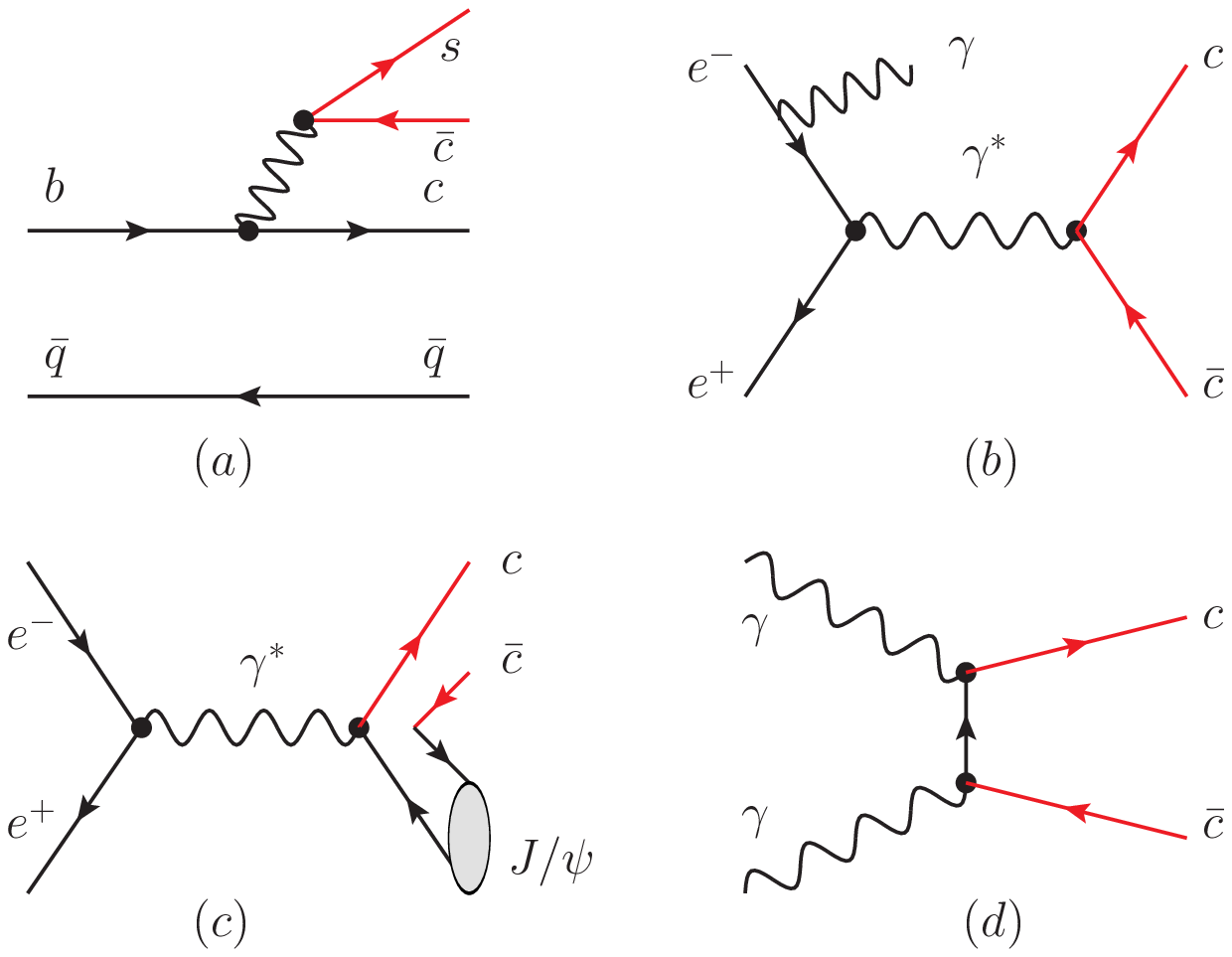}}
\end{tabular}
\caption{(color online). The diagramatic description of the production mechanism of $XYZ$ states. Here,
the $B$ meson decay ($B\to K+XYZ$), $e^+e^-$ annihilation ($e^+e^-\to XYZ$), the double charm production ($e^+e^-\to J/\psi+XYZ$), the $\gamma\gamma$ fusion process ($\gamma\gamma\to XYZ$)
correspond to diagrams (a)-(d), respectively.
\label{pr} }
\end{figure}
\end{center}

Since the observation of $X(3872)$ in 2003, more and more charmoniumlike states referred to $XYZ$ have been announced by experiments after analyzing various processes.
Until now, the family of $XYZ$ states has increasingly become abundant and the number of the states reaches 23. In general, the observed $XYZ$ states can be categorized into five groups, which correspond to five different production mechanisms, i.e., the $B$ meson decay ($B\to K+XYZ$), $e^+e^-$ annihilation ($e^+e^-\to XYZ$), the double charm production ($e^+e^-\to J/\psi+XYZ$), the $\gamma\gamma$ fusion process ($\gamma\gamma\to XYZ$), and the hidden-charm/bottom dipion and open-charm decays of higher charmonia/bottomonia and charmoniumlike/bottomoniumlike states (see Fig. \ref{pr}). According to the above classification, we list all the reported $XYZ$ states in Table \ref{xyz}.

These newly observed $XYZ$ states provide us a good platform to study the nonperturbative behavior of QCD, which is one of the reasons why these experimental observations arouse theorists' extensive interest. The importance of the study of the $XYZ$ states is also due to the possibility that the observed $XYZ$ states can be the potential candidates of exotic states. In the past decade, theorists have also paid great attention to $XYZ$ states and have made a big progress on revealing the underlying mechanisms behind these novel phenomena. Thus, in this review paper, we briefly summarize the present experimental and theoretical status of the study of $XYZ$.

\renewcommand{\arraystretch}{1.2}
\begin{table}[htbp]
\caption{A summary of the observed $XYZ$ states. Here, we use A, B, C, D, and E to mark the processes,
$B$ meson decay, $e^+e^-$ annihilation, the double charm production, $\gamma\gamma$ fusion process, and the hidden-charm/bottom dipion and open-charm/bottom decays of higher charmonia/bottomonia and charmoniumlike/bottomoniumlike states, respectively.
\label{xyz}}
\begin{tabular}{ccccccccc}
\toprule[1pt]
A \cite{Choi:2003ue,Abe:2004zs,Choi:2007wga,Mizuk:2008me,Aaltonen:2009tz}&B \cite{Aubert:2005rm,Yuan:2007sj,Aubert:2006ge,Wang:2007ea,Pakhlova:2008vn}&C \cite{Abe:2007jna,Abe:2007sya}&D \cite{Uehara:2005qd,Shen:2009vs,Uehara:2009tx}&E \cite{Belle:2011aa,Ablikim:2013mio,Ablikim:2013emm,Ablikim:2013wzq,Ablikim:2013xfr}\\\midrule[1pt]
$X(3872)$ &$Y(4260)$&$X(3940)$&$X(3915)$&$Z_b(10610)$\\
$Y(3940)$&$Y(4008)$&$X(4160)$&$X(4350)$&$Z_b(10650)$\\
$Z^+(4430)$&$Y(4360)$&--&$Z(3930)$&$Z_c(3900)$\\
$Z^+(4051)$&$Y(4660)$&--&--&$Z_c(4025)$\\
$Z^+(4248)$&$Y(4630)$&--&--&$Z_c(4020)$\\
$Y(4140)$&--&--&--&$Z_c(3885)$\\
$Y(4274)$&--&--&--&--\\
\bottomrule[1pt]
\end{tabular}
\end{table}

This review paper is organized as follows. After introduction, we review the experimental and theoretical progress on $XYZ$ states in Sects. \ref{sec2}-\ref{sec6}, which are produced from $B$ meson decay, $e^+e^-$ annihilation, the double charm production, $\gamma\gamma$ fusion process, and the hidden-charm/bottom dipion and open-charm/bottom decays of higher charmonia/bottomonia and charmoniumlike/bottomoniumlike states, respectively. The last section is devoted to the conclusion.

\section{The $XYZ$ states from B meson decays}\label{sec2}

As shown in Table \ref{xyz}, the $B$ meson decay is a suitable platform to produce $XYZ$ states.
Until now, experiments have reported seven $XYZ$ states. The chains of the production and decays of $X(3872)$, $Y(3940)$, $Z^+(4430)$, $Z^+(4051)$, $Z^+(4248)$, $Y(4140)$, and $Y(4274)$ \cite{Choi:2003ue,Abe:2004zs,Choi:2007wga,Mizuk:2008me,Aaltonen:2009tz} summarized as follows
\begin{eqnarray}
B\to \left\{\begin{array}{l}
\begin{array}{l}
X(3872) K\to \underline{J/\psi \pi^+\pi^-} K
\end{array}\\
\begin{array}{l}
Y(3940)K\to \underline{J/\psi\omega} K
\end{array}\\
\begin{array}{l}
Z^+(4430)K\to \underline{\psi^\prime \pi^+} K
\end{array}\\
\left.
\begin{array}{l}
Z^+(4051)K\\
Z^+(4248)K\\
\end{array}\right\}
\to\underline{\chi_{c1} \pi^+} K\\
\left.
\begin{array}{l}
Y(4140)K\\
Y(4274)K\\
\end{array}\right\}\to \underline{J/\psi\phi} K\\
\end{array}\right. , \label{a1}
\end{eqnarray}
where we have used underlines to denote the corresponding decay channels. We need to emphasize that we only list one typical decay channel for $X(3872)$. Below we will present more detailed description of the experimental status of $X(3872)$.

\subsection{$X(3872)$}

\begin{center}
\begin{figure}[htb]
\begin{tabular}{cccc}
\scalebox{0.46}{\includegraphics{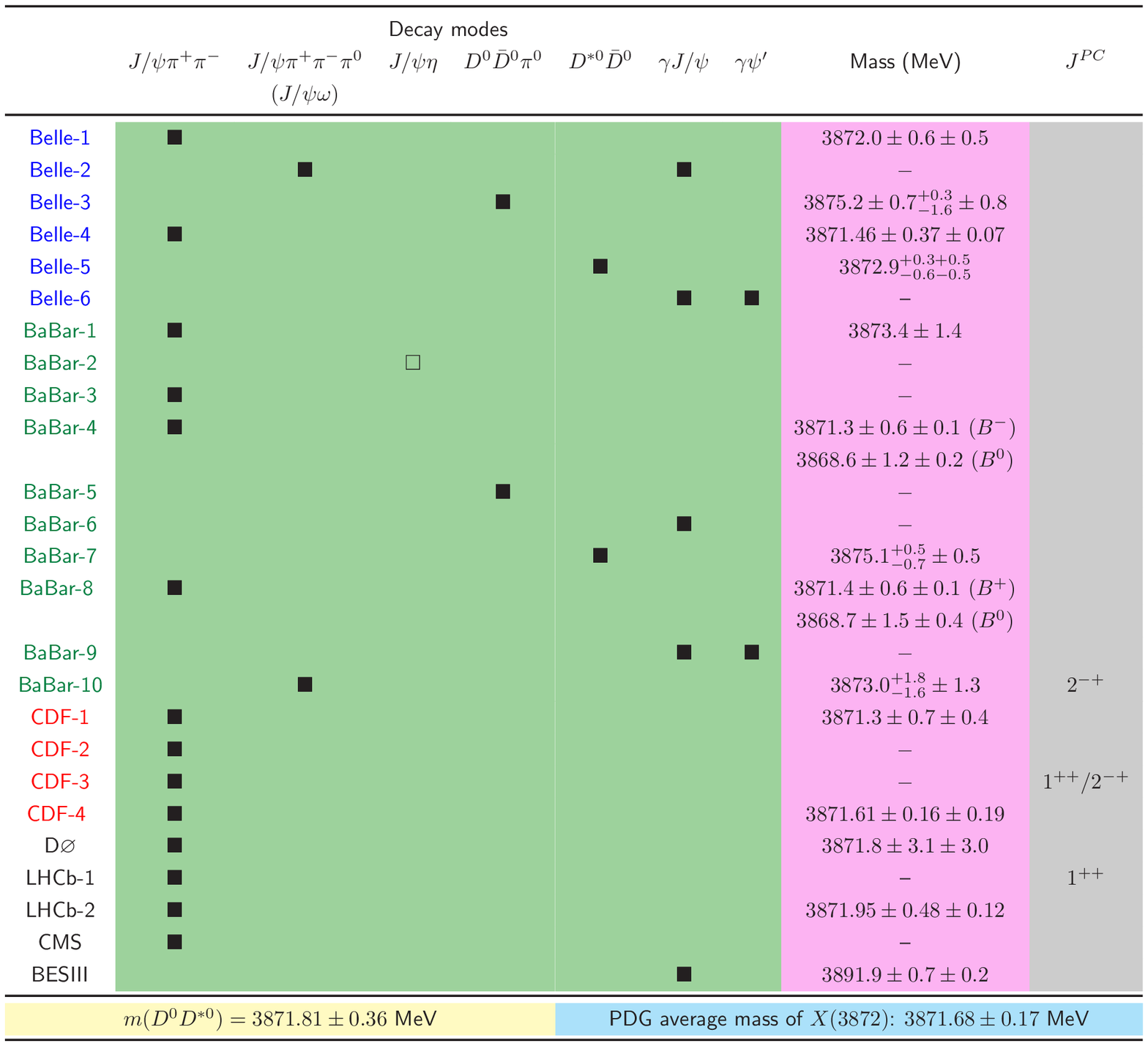}}
\end{tabular}
\caption{(color online). The experimental measurements of $X(3872)$ for different experiments.
The experimental information is from Refs. \cite{Choi:2003ue,Abe:2005ix,Gokhroo:2006bt,:2008te,:2008su,Bhardwaj:2011dj}, which are marked by {\sf Belle-i (i=1-6)}, respectively. The experimental results in Refs. \cite{Aubert:2004ns,Aubert:2004fc,Aubert:2005eg,Aubert:2005zh,Aubert:2005vi,Aubert:2006aj,Aubert:2007rva,
Aubert:2008gu,:2008rn,delAmoSanchez:2010jr} are marked by {\sf BaBar-j (j=1-10)}, respectively. The CDF results are marked by {\sf CDF-k (k=1-4)}, which correspond to Refs. \cite{Acosta:2003zx,Abulencia:2005zc,Abulencia:2006ma,Aaltonen:2009vj}, respectively. The D$\varnothing$ result is taken from Ref. \cite{Abazov:2004kp}. Recently, LCHb \cite{Aaij:2013zoa,Aaij:2011sn}, CMS \cite{Chatrchyan:2013cld} and BESIII \cite{Ablikim:2013dyn} also studied $X(3872)$, where we use LCHb-1 and LHCb-2 to distinguish the results in Refs. \cite{Aaij:2013zoa,Aaij:2011sn}, respectively. Here, we also list the average mass of $X(3872)$ and the threshold of $D^0D^{*0}$ given by the particle data group (PDG) \cite{Beringer:1900zz}. $\blacksquare$ and $\square$ denote observed and unobserved decay channels indicated in experiments, respectively. The $B^{\pm,0}$ in the bracket denotes the measured mass coming from the $B^{\pm,0}\to X(3872)K^{\pm,0}$ decay process.
\label{mass3872} }
\end{figure}
\end{center}

In 2003, the Belle Collaboration first reported the observation of $X(3872)$ in the $J/\psi\pi^+\pi^-$ invariant mass spectrum of $B\to K J/\psi\pi^+\pi^-$ \cite{Choi:2003ue}. $X(3872)$ is the first observed charmoniumlike state and it should be noted that the experimental information of $X(3872)$ is the most abundant among all the observed $XYZ$ states. CDF, D$\varnothing$, BaBar, LHCb, and CMS have later confirmed $X(3872)$ with the observations of more decay channels of $X(3872)$. As listed in the particle data group (PDG), there exist the different experimental values of the $X(3872)$ mass for different experiments. In the following, we further summarize the experimental status of $X(3872)$, which is shown in Fig. \ref{mass3872}.

According to the quark model calculation, the mass of $2^3P_1$ charmonium ($\chi_{c1}^\prime$) is not consistent with that of $X(3872)$, where the mass difference between $\chi_{c1}^\prime$ and $X(3872)$ reaches $50\sim200$ MeV. In addition, an isospin scalar charmonium into $J/\psi\rho$ is a typical isospin violating decay. Due to the above difficulty of $X(3872)$ as $\chi_{c1}^\prime$, different theoretical explanations for $X(3872)$ were proposed, which include the molecular state
\cite{Close:2003sg,Voloshin:2003nt,Wong:2003xk,Swanson:2003tb,Tornqvist:2004qy},
the $1^{++}$ cusp \cite{Bugg:2004rk}, the S-wave threshold effect due
to the $D^0\bar{D}^{0*}$ threshold \cite{Rosner:2006vc}, the hybrid
charmonium \cite{Li:2004sta}, the diquark anti-diquark bound state
\cite{Maiani:2004vq} and the tetraquark state \cite{3872-tetra,Cui:2006mp}.

Among these theoretical proposals to the structure of $X(3872)$, the molecular state explanation is the most popular one. Up to now, there have been several groups performing the dynamical study of the
molecular assignment of $X(3872)$. Swanson once suggested that $X(3872)$
was a $D^0 \bar{D}^{*0}$ molecular state bound by both the pion
exchange and quark exchange \cite{Swanson:2003tb}. Following the method proposed by
T\"{o}rnqvist \cite{Tornqvist:1993vu,Tornqvist:1993ng}, the
potential between $D^0 \bar{D}^{*0}$ through exchanging a single
pion was obtained, where the formalism is based on a microscopic
quark-pion interaction. Swanson indicated that one pion exchange
alone can not bind $D$ and $\bar D^*$ to form a molecule. He also introduced the short-range
quark-gluon force \cite{Swanson:2003tb}.
Wong calculated the $D\bar{D}^*$ system in the
quark model containing a four-body non-relativistic Hamiltonian
with pairwise effective interactions \cite{Wong:2003xk}, which is similar to
the consideration of adding short-range quark-gluon force in
Ref. \cite{Swanson:2003tb}. Here, an S-wave
$D\bar{D}^*$ molecule was found with the binding energy $\sim 7.53$ MeV. Further investigations based on the
molecular assumption are later performed in Refs.
\cite{AlFiky:2005jd,Fleming:2007rp,Braaten:2007ct,Hanhart:2007yq,Voloshin:2007hh}.

There are different conclusions of whether $X(3872)$ is a $D^0 \bar{D}^{*0}$ molecular state.
Suzuki obtained the one pion exchange potential (OPEP) by using the
effective Lagrangian arguing that $X(3872)$ is not a
molecular state of $D^0\bar{D}^{*0}+\bar{D}^0 D^{*0}$
\cite{Suzuki:2005ha}. He also emphasized that introducing the short-range quark-gluon force
is not suitable for studying the $D\bar{D}^*$ molecular system.

To further clarify the underlying property of $X(3872)$, more theoretical groups joined the discussion of $X(3872)$ and the study of the interaction between $D$ and $\bar{D}^*$. The lesson from studying the deuteron is that the one pion
exchange potential alone does not bind the proton and neutron pair
into the deuteron in nuclear physics. In fact, the strong
attractive force in the intermediate range has to be introduced in
order to bind the deuteron, which is elegantly modeled by the sigma meson
exchange potential. Thus, in Ref. \cite{Liu:2008fh}
the authors performed a dynamical calculation of the $D^0\bar{D}^{*0}$ system by considering the pion and sigma meson exchange potential. The result disfavors the interpretation of $X(3872)$ as a loosely bound molecular state if we use the experimental $D^*D\pi$ coupling constant $g=0.59$  and a reasonable cutoff around 1 GeV, which is the typical hadronic scale \cite{Liu:2008fh}. Later, Thomas and Close confirmed the above results and indicated that charged modes of $D\bar{D}$ is important \cite{Thomas:2008ja}. In Ref. \cite{Lee:2009hy} Lee {\it et al.} also discussed the possibility of $X(3872)$ as a hadronic $D\bar{D}^*$ molecular state, where the pseudoscalar, scalar and vector meson exchanges are included and the isospin symmetry breaking effect is also considered. They found the bound state solution of $D\bar{D}^*$ system with $J^{PC}=1^{++}$ \cite{Lee:2009hy}. Li and Zhu further studied $X(3872)$ as a $D\bar{D}^*$ molecular state by the one-pion-exchange (OPE) model and the one-boson-exchange (OBE) model. They took into account the S-D wave mixing, the mass difference between the neutral and charged $D$($D^*$) mesons and the coupling of the $D$($D^*$) pair to $D^*\bar{D}^*$. $X(3872)$ can be quite naturally explained as a loosely bound molecular state \cite{Li:2012cs}.

The molecular picture
naturally explains both the proximity of $X(3872)$ to the
$D^0\bar{D}^{*0}$ threshold and the isospin violating $J/\psi
\rho$ decay mode. It predicted the decay width of the
$J/\psi\pi^+\pi^-\pi^0$ mode to be comparable with that of
$J/\psi\rho$, which was confirmed by Belle collaboration
\cite{Abe:2005ix}. Within the same picture, Brateen and Kusunoki
predicted that the branching ratio of $B^0\to X(3872)K^0$ is
suppressed by more than one order of magnitude compared to that of
$B^+\to X(3872)K^+$ \cite{Braaten:2004ai}.

Both the Belle and Babar collaborations observed the radiative
decay mode. The Belle's measurement gives \cite{Abe:2005ix}
\begin{equation}
\frac{BR[X(3872)\to \gamma J/\psi]}{BR[X(3872)\to J/\psi
\pi^+\pi^-]}=0.14\pm 0.05
\end{equation}
while the Babar Collaboration obtains \cite{Aubert:2006aj}
\begin{equation}
\frac{BR[X(3872)\to \gamma J/\psi]}{BR[X(3872)\to J/\psi
\pi^+\pi^-]}\approx0.25 \; ,
\end{equation}
which are against the prediction by the molecular picture
$7\times 10^{-3}$.

In addition, the Belle Collaboration measured the ratio
\cite{Gokhroo:2006bt}
\begin{equation}
\frac{BR[X(3872)\to D^0\bar{D}^0\pi^0]}{BR[X(3872)\to \pi^+\pi^-
J/\psi]}=9.4^{+3.6}_{-4.3}
\end{equation}
which is much larger than the theoretical value $0.054$ due to the
molecular assumption. From Ref. \cite{Gokhroo:2006bt}, one can
also extract
\begin{equation}
\frac{BR[B^0\to X(3872)K^0]}{BR[B^+\to X(3872)K^+]}\approx 1.62
\end{equation}
which is also much larger than the molecule prediction.

Instead,
$X(3872)$ may have a dominant $c\bar{c}$ component with some
admixture of $D^0\bar{D}^{*0}+\bar{D}^0 D^{*0}$
\cite{Meng:2005er,Suzuki:2005ha}.  In the following, we need to introduce several
studies using the coupled-channel model. Kalashnikova indicated that the coupling of the bare $2^3P_1$ state to $D\bar{D}^*$ channel can generate a near-threshold virtual state with the energy of about 0.3 MeV, which can correspond to $X(3872)$ \cite{Kalashnikova:2005ui}. In Ref. \cite{Danilkin:2010cc}, the authors
indicated that the mass and width of $X(3872)$ can be well explained by their dynamical mechanism, and emphasize that their result partly supports $X(3872)$ as an ordinary $2^3P_1$ state of $c\bar{c}$ origin, which is concluded  in Ref. \cite{Zhang:2009bv}.
Recently, Coito, Rupp and Beveren pointed out that $X(3872)$ is not a genuine meson-meson molecule due to the mixing with the corresponding quark-antiquark states \cite{Coito:2012vf,Coito:2010if}.
In addition, the ratio
\begin{eqnarray}
\frac{BR(X(3872)\to \psi^\prime\gamma)}{BR(X(3872)\to J/\psi\gamma)}=3.4\pm 1.4
\end{eqnarray}
was given by BaBar \cite{:2008rn}, which is not consistent with the prediction under the explanation of the $D\bar{D}^*$ molecular state \cite{Swanson:2004pp}.

In the following, we need to introduce a lattice simulation of the study of $X(3872)$. In Ref. \cite{Prelovsek:2013cra}, authors performed the dynamical $N_f=2$ lattice simulation with $J^{PC}=1^{++}$ and $I=0$, which shows that
there exists a candidate for $X(3872)$ below the $DD^*$ threshold. In addition, they also obtained large and negative $DD^*$ scattering length $a_0=-1.7 \pm 0.4$ fm and the effective range $r_0=0.5 \pm 0.1$ fm.

Before closing this subsection, we need to give a comment to the tetraquark explanation of $X(3872)$. In Ref. \cite{Maiani:2004vq}, Maiani {\it et al.} predicted the tetraquark states $(cu)(\bar{c}\bar{u})$, $(cd)(\bar{c}\bar{u})$ and $(cd)(\bar{c}\bar{d})$. However, the BaBar Collaboration indicated that there no evidence of a charged partner of $X(3872)$ by studying $B\to J/\psi\pi^-\pi^0$ \cite{Aubert:2004zr}.
Thus, the tetraquark explanation \cite{Maiani:2004vq} for $X(3872)$ can be excluded.

\subsection{$Y(3940)$, $Y(4140)$ and $Y(4274)$}

The CDF Collaboration announced a new charmonium-like state $Y(4140)$ by analyzing the
$J/\psi\phi$ invariant mass spectrum in $B\to K J/\psi \phi$ channel, which results in the $C$-parity and $G$-parity of $Y(4140)$ being even. The measured mass and width
of $Y(4140)$ are $4143.0\pm2.9(\mathrm{stat})\pm1.2(\mathrm{syst})$ MeV and $11.7^{+8.3}_{-5.0}(\mathrm{stat})\pm 3.7(\mathrm{syst})$ MeV \cite{Aaltonen:2009tz}, respectively.

By comparing $Y(4140)$ with a series of charmonium-like states $X$, $Y$, and $Z$, one notices that $Y(4140)$ is similar to $Y(3940)$, which is a charmonium state with
$m=3943\pm11(\mathrm{stat})\pm13(\mathrm{syst})$ MeV and $\Gamma=
87\pm 22(\mathrm{stat})\pm 26(\mathrm{syst})$ MeV reported by the Belle Collaboration \cite{Abe:2004zs} and confirmed by the Babar Collaboration \cite{Aubert:2007vj}.
Both $Y(4140)$ and $Y(3940)$ were observed in
the mass spectrum of $J/\psi+{light\, vector\, meson}$ in the $B$ meson decay
\begin{eqnarray*}
B\to K+\Bigg\{\begin{array}{cc} \underline{J/\psi \phi} & \Longrightarrow Y(4140)\\
\underline{J/\psi \omega} &\Longrightarrow Y(3940)\end{array}.\label{channel}
\end{eqnarray*}
The mass difference between $Y(4140)$ and $Y(3940)$ is approximately equal to that between $\phi$ and $\omega$ mesons:
$$M_{Y(4140)}-M_{Y(3930)}\sim M_\phi-M_\omega.$$
Additionally, $Y(4140)$ and $Y(3940)$ are close to the thresholds of $D_s^*\bar{D}_s^*$ and $D^*\bar{D}^*$ respectively, and satisfy an almost exact mass relation
\begin{eqnarray*}
M_{Y(4140)}-2M_{D_s^*}\approx M_{Y(3940)}-2M_{D^*}.\label{relation}
\end{eqnarray*}

The above similarities provoke a uniform molecular picture to reveal the underlying structure of $Y(4140)$ and $Y(3940)$ \cite{Liu:2009ei,Liu:2008tn}. The flavor wave functions of $Y(4140)$ and $Y(3940)$
are \cite{Liu:2009ei,Liu:2008tn}
\begin{eqnarray*}
|Y(4140)\rangle&=&|D_s^{*+}D_s^{*-}\rangle,\\
|Y(3940)\rangle&=&\frac{1}{\sqrt{2}}\Big[|D^{*0}\bar{D}^{*0}\rangle+|D^{*+}D^{*-}\rangle\Big].
\end{eqnarray*}
A selection rule for the quantum numbers of $Y(4140)$ and $Y(3940)$ is observed under the
$D_{s}^*\bar{D}_{s}^*$ and $D^*\bar{D}^*$ molecular state assignments, respectively.
The possible quantum numbers of the S-wave
vector-vector system are $J^{P}=0^+, 1^+, 2^+$. However for the
neutral $D^\ast {\bar D}^\ast$ system with $C=+$, we can have
$J^{P}=0^+$ and $2^+$ only since $C=(-1)^{L+S}$ and $J=S$ with
$L=0$ \cite{Liu:2009ei}, which provides important criterion to test molecular state explanation for $Y(3940)$ and $Y(4140)$.

To answer whether $D^*\bar{D}^*$ or $D_s^*\bar{D}_s^*$ system can be bound, a dynamical calculation was performed in Ref. \cite{Liu:2008tn} by the effective Lagrangian approach. Here, the exchanged mesons between $D^*\bar{D}^*$ ($D_{s}^*\bar{D}^*_s$) include the pseudoscalar, vector and
$\sigma$ mesons (see Ref. \cite{Liu:2008tn} for the details of the derivation of the exchange
potential). For $Y(4140)$ and $Y(3940)$ states with $J^P=0^+,\,2^+$, the molecular solution has been found. Later, the study in Refs. \cite{Mahajan:2009pj,Branz:2009yt,Albuquerque:2009ak,Ding:2009vd,Zhang:2009st} further supports the molecular explanation for $Y(4140)$ and $Y(3940)$.

Besides the dynamical study of $Y(4140)$ and $Y(3940)$, it is an important research topic to investigate the decay behavior of $Y(4140)$ and $Y(3940)$, which includes the hidden-charm decay, the open-charm decay, radiative decay and double-photon decay. In Ref. \cite{Liu:2009iw}, we study the hidden-charm decay of $Y(4140)$ assuming $Y(4140)$ as the second radial excitation of the P-wave charmonium $\chi_{cJ}^{\prime\prime}$ ($J = 0, 1$). The result indicates that the upper
limit of the branching ratio of the hidden charm decay $Y(4140)\to
J/\psi\phi$ is of the order of $10^{-4}\sim 10^{-3}$ for both of
the charmonium assumptions for $Y(4140)$, which disfavors the
large hidden charm decay pattern indicated by the CDF experiment.
This means that the assumption of the pure second radial excitation of the
P-wave charmonium $\chi_{cJ}^{\prime\prime}$ ($J=0,\,1$) for $Y(4140)$ is
problematic \cite{Liu:2009iw}.

As indicated in Ref. \cite{Liu:2009ei}, the line shapes of the photon spectrum of $Y(4140)\to { D}_s^{\ast+} D_s^- \gamma$ and $Y(3940)\to{D}^{\ast+} D^-\gamma$ are crucial to test the molecular state assignment to $Y(4140)$ and $Y(3940)$. A calculation of the radiative decay of $Y(4140)$ and $Y(3930)$ was later performed \cite{Liu:2009pu}. According to the results of the photon spectrum in $Y(4140)\rightarrow
D_s^{*+}D_s^-\gamma$ and $Y(3940)\rightarrow D^{*+}D^-\gamma$, we suggest further experimental study on the radiative decay of $Y(4140)$ and $Y(3940)$.

By checking the CDF data \cite{Aaltonen:2009tz}, we also notice that there exists another enhancement structure around 4270 MeV besides the $Y(4140)$ signal in the $J/\psi \phi$ mass spectrum of $B^+\to J/\psi \phi K^+$, which has lower significance than that of $Y(4140)$. CDF later reported a new structure $Y(4274)$ in the $J/\psi\phi$ invariant mass spectrum \cite{Yi:2010aa}. In Ref. \cite{Liu:2010hf}, the explanation of the S-wave $D_s\bar{D}_{s0}(2317)$ molecular state for $Y(4274)$ was proposed and the S-wave $D\bar{D}_0(2400)$ molecular state was predicted, which is the partner of $Y(4274)$. A calculation by the QCD sum rule also supports the above proposal \cite{Finazzo:2011he}. In addition, the open-charm radiative and pionic decays of $Y(4274)$ were obtained
in Ref. \cite{He:2011ed}.

Finally, we need to introduce the recent experimental progress of $Y(4140)$.
After the observation of $Y(4140)$ given by CDF \cite{Aaltonen:2009tz}, the LHCb
Collaboration indicated that no evidence for $Y(4140)$ is found by carrying out the search for $Y(4140)$ in $B^+\to J/\psi\phi K^+$ \cite{Aaij:2012pz}. However, very recently the D$\varnothing$ Collaboration \cite{Abazov:2013xda} and the CMS Collaboration \cite{Chatrchyan:2013dma} confirmed the observation of $Y(4140)$. Besides the above observations, D$\varnothing$ also reported a second enhancement ar a mass of $4328.5\pm12.0$ MeV \cite{Abazov:2013xda}, while CMS also found the evidence of an additional enhancement with mass $M=4313.8\pm5.3\pm7.3$ MeV and width $\Gamma=38^{+30}_{-15}\pm16$ MeV \cite{Chatrchyan:2013dma}.

\subsection{$Z^+(4430)$, $Z^+(4051)$ and $Z^+(4248)$}

As a charged charmonium-like state, $Z^+(4430)$ was observed by Belle with measured mass $m=4433\pm4\pm2$ MeV and width $\Gamma=45^{+18+30}_{-13-13}$ \cite{Choi:2007wga}. However, $Z^+(4430)$ was not confirmed by BaBar \cite{Aubert:2008aa}.

Different theoretical explanations to $Z^+(4430)$ were given, which include
S-wave threshold effect of $D_1(2420)\bar{D}^*(2010)$ \cite{Rosner:2007mu}, $D_1(2420)\bar{D}^*(2010)$ molecular state \cite{Meng:2007fu}, tetraquark state \cite{Maiani:2007wz}, cusp effect \cite{Bugg:2007vp}, $\Lambda_c\Sigma_c^0$ bound state \cite{Qiao:2007ce}. In Ref. \cite{Cheung:2007wf}, the authors predicted the bottomed analog of $Z^+(4430)$ if $Z^+(4430)$ is $(cu)(\bar c \bar d)$ tetraquark state. The QCD sum rule study of $Z^+(4430)$ indicates that $Z^+(4430)$ can be a $D^*\bar{D}_1$ molecule with $J^P=0^-$ \cite{Lee:2008gn}. Braaten and Lu studied the line shape of $Z^+(4430)$ \cite{Braaten:2007xw}.

In Refs. \cite{Liu:2007bf,Liu:2008xz}, the authors investigate whether $Z^+(4430)$ is a loosely bound S-wave state of $D^*\bar{D}_1$ or $D^*\bar{D}_1^\prime$ with $J^P=0^-,1^-,2^-$. They notice that the attraction by the one pion exchange potential alone is not strong enough to form a bound state with realistic pionic coupling constants deduced from the decay widths of $D_1$ and $D^\prime_1$. If considering both pion and sigma meson exchange potentials, they found that the S-wave $D_1\bar{D}^*$   molecular state with only $J^P=0^-$ and $D^\prime_1\bar{D}^*$ molecular states with $J^P=0^-, 1^-, 2^-$   may exist with reasonable parameters \cite{Liu:2008xz}.

Besides $Z^+(4430)$, two charged charmoniumlike states $Z^+(4051)$ and $Z^+(4248)$
were later reported by Belle \cite{Mizuk:2008me}, which was not confirmed by BaBar \cite{Lees:2011ik}.
The masses and widths of $Z^+(4051)$ and $Z^+(4248)$ are \cite{Mizuk:2008me}
\begin{eqnarray*}
M_{Z^+(4051)}&=&4051\pm14^{+20}_{-41}\, \mathrm{MeV},\\
M_{Z^+(4248)}&=&4248^{+44+180}_{-29-35}\, \mathrm{MeV},\\
\Gamma_{Z^+(4051)}&=&82^{+21+47}_{-17-22}\, \mathrm{MeV},\\
\Gamma_{Z^+(4248)}&=&177^{+54+316}_{-39-61}\, \mathrm{MeV}.
\end{eqnarray*}
Since the mass of $Z^+(4051)$ is slightly above the $D^*\bar{D}^*$ threshold, it is possible to assume $Z^+(4051)$ as a $D^*\bar{D}^*$ molecular state. However, a dynamical study of $D^*\bar{D}^*$ molecular state shows that there exist bound state solutions for the $J^P=0^+, 1^+$ $D^*\bar{D}^*$ systems only with large cutoff \cite{Liu:2008tn}. Later, by solving the resonating group method equation in the chiral SU(3) quark model, Liu and Zhang indicated that $Z^+(4051)$ is unlikely to be an S-wave $D^*\bar{D}^*$ molecule \cite{Liu:2008mi}.

As for $Z^+(4248)$, its mass is near the $D_1\bar{D}$ or $D_0\bar{D}^*$ threshold \cite{Mizuk:2008me}.
Thus, Ding studied the possibility of $Z^+(4248)$ as a hadronic molecular state and found that
$Z^+(4248)$ disfavors the $D_1\bar{D}$ or $D_0\bar{D}^*$ molecular state.

At present, only Belle reported $Z^+(4430)$, $Z^+(4051)$ and $Z^+(4248)$. Further confirmation and experimental study of these three charged charmoniumlike states in other experiments is still an important topic.

\section{$Y$ states directly from the $e^+e^-$ annihilation}\label{sec3}

The $e^+e^-$ annihilation is also an ideal  process to produce $XYZ$ states. As shown in Table \ref{xyz}, experiments have observed five $Y$ states, which have $J^{PC}=1^{--}$ quantum number. Among these states, only $Y(4008)$ announced by the Belle Collaboration \cite{Yuan:2007sj} has not been confirmed by other experiments. At present, the hidden-charm dipion decays of $Y(4260)$ \cite{Aubert:2005rm}, $Y(4008)$
\cite{Yuan:2007sj}, $Y(4360)$ \cite{Aubert:2006ge}, and $Y(4660)$ \cite{Wang:2007ea} were experimentally observed while $Y(4630)$ \cite{Pakhlova:2008vn} has open-charm decay mode, i.e.,
\begin{eqnarray}
e^+e^-&&\to \left\{\begin{array}{ll}
\left.
\begin{array}{l}
{Y(4260)}\to \underline{J/\psi\pi^+\pi^-}
\end{array}\right.\\
\left.
\begin{array}{l}
{Y(4008)}\\
{Y(4360)} \\
{Y(4660)} \\
\end{array} \right\}\to\underline{\psi^\prime \pi^+\pi^-}
\\
\left.
\begin{array}{l}
{Y(4630)}\to \underline{\Lambda_c\bar{\Lambda}_c}
\end{array}\right.
\end{array}\right. .
\end{eqnarray}

\subsection{$Y(4260)$ and $Y(4008)$}

$Y(4260)$ was observed by BaBar in $e^+e^-\to J/\psi\pi^+\pi^-$ \cite{Aubert:2005rm}. Later, Belle also confirmed $Y(4260)$ by the same process, and indicated that there is another enhancement structure $Y(4008)$ \cite{Yuan:2007sj}.
In Table. \ref{exp}, we summarize the information of resonance parameters of $Y(4260)$ from different experiments.
\renewcommand{\arraystretch}{1.6}
\begin{table}[htb]
\caption{The experimental information of $Y(4260)$.   \label{exp}}
\begin{center}
\begin{tabular}{c|cc} \toprule[1pt]
  Experiment                                & Mass (MeV)                        & Width (MeV)  \\\midrule[1pt]
  BaBar \cite{Aubert:2005rm}        &$4259\pm8^{+2}_{-6}$               &$88\pm23^{+6}_{-4}$\\
  CLEO \cite{He:2006kg}             &$4284^{+17}_{-16}\pm4$             &$73^{+39}_{-25}\pm5$ \\
  Belle \cite{Abe:2006hf}           &$4295\pm10^{+10}_{-3}$             &$133\pm26^{+13}_{-6}$\\
  Belle \cite{Yuan:2007sj}          &$4247\pm12^{+17}_{-32}$            &$108\pm19\pm10$\\
  BaBar \cite{Aubert:2008aj}        &$4252\pm6^{+2}_{-3}$               &$105\pm18^{+4}_{-6}$\\
  BaBar \cite{Lees:2012cn}          &$4245\pm5\pm4$                     &$114^{+16}_{-15}\pm7$\\
  Belle \cite{Liu:2013dau}          &$4258.6\pm8.3\pm12.1$              &$134.1\pm16.4\pm5.5$\\     
\bottomrule[1pt]
\end{tabular}
\end{center}
\end{table}

The observation of $Y(4260)$ has stimulated extensive discussions of its structure. There are two main opinions, i.e., either explaining it as an exotic state or categorizing it into the conventional charmonium family.

After the observation of $Y(4260)$, different exotic state explanations were proposed, which mainly include charmonium hybrid \cite{Zhu:2005hp,Kou:2005gt,Close:2005iz}, diquark-antidiquark state $[cs][\bar{c}\bar{s}]$ \cite{Maiani:2005pe,Ebert:2008wm}, different molecular state assignments \cite{Liu:2005ay,Yuan:2005dr,Qiao:2005av,Ding:2008gr,MartinezTorres:2009xb,Close:2010wq}, and charmonium hybrid state with strong coupling with $D\bar{D}_1$ and $D\bar{D}_0$ channels \cite{Kalashnikova:2008qr}. Although there are these exotic state possibilities for $Y(4260)$, the lack of the signal of $Y(4260)$ in certain channels also poses a serious question to these exotic state explanations mentioned above.

Theorists tried to explain $Y(4260)$ to be the conventional charmonium. In Ref. \cite{LlanesEstrada:2005hz}, the mixing of $4S$ and $3D$ vector charmonia was suggested for $Y(4260)$. Eichten and Quigg calculated the decay behavior of $2^3D_1$ $c\bar{c}$ state and excluded this assignment to $Y(4260)$ \cite{Eichten:2005ga}. By analyzing the mass spectrum, the authors in Ref. \cite{Segovia:2008zz} indicated that $Y(4260)$ cannot be categorized into the charmonium family. However, Li and Chao calculated the mass spectrum of charmonium with the screened potential \cite{Li:2009zu}. The obtained mass of $\psi(4S)$ is close to that of $Y(4260)$. Thus, $Y(4260)$ as a $\psi(4S)$ state cannot be excluded.
If explaining $Y(4260)$ as a conventional $c\bar{c}$ state, the main challenge is that there is no evidence of $Y(4260)$ in the obtained open-charm processes \cite{Abe:2006fj,Pakhlova:2008zza,Pakhlova:2007fq,Pakhlova:2009jv} and $R$-value scan \cite{Burmester:1976mn,Brandelik:1978ei,Siegrist:1981zp,Osterheld:1986hw,Bai:1999pk,Bai:2001ct,CroninHennessy:2008yi,Ablikim:2009ad}.

Later, the non-resonant explanation to $Y(4260)$ was proposed in Ref. \cite{Chen:2010nv}. By the interference of $e^+e^-\to \psi(4160)/\psi(4415)\to J/\psi\pi^+\pi^-$ and the background contribution to
$e^+e^-\to J/\psi\pi^+\pi^-$, the $Y(4260)$ structure can be well reproduced \cite{Chen:2010nv}. This non-resonant explanation to $Y(4260)$ can answer why there is no evidence of $Y(4260)$ in the exclusive open-charm decay channels \cite{Abe:2006fj,Pakhlova:2008zza,Pakhlova:2007fq,Pakhlova:2009jv} and the $R$-value scan \cite{Burmester:1976mn,Brandelik:1978ei,Siegrist:1981zp,Osterheld:1986hw,Bai:1999pk,Bai:2001ct,CroninHennessy:2008yi,Ablikim:2009ad} mentioned above.

Besides confirming the observation of $Y(4260)$, Belle also reported a broad structure $Y(4008)$ in the $J/\psi\pi^+\pi^-$ invariant mass spectrum. In Ref. \cite{Liu:2007ez}, the author discussed the possible assignments for this enhancement which include $\psi(3S)$ and $D^*\bar{D}^*$ molecular state. Here, the hidden-charm and open-charm decays were studied, which will be helpful to distinguish two different assignments to $Y(4008)$. Ding studied the $D^*\bar{D}^*$ interaction and found that the $D^*\bar{D}^*$ molecular state with $J^{PC}=1^{--}$ can exist. However, if this $D^*\bar{D}^*$ molecular state corresponds to $Y(4008)$, we must explain why $Y(4008)$ is very broad \cite{Ding:2009zq}.
We also notice an interesting phenomenon in Ref. \cite{Chen:2010nv}, where the $Y(4008)$ structure
can be reproduced by the interference effect.

Recently there were several recent experimental progresses relevant to the hidden-charm dipion, open-charm, and radiative decays of $Y(4260)$.
In 2013, several charged charmoniumlike states  $Z_c(3900)$ \cite{Ablikim:2013mio}, $Z_c(4025)$ \cite{Ablikim:2013emm}, $Z_c(4020)$ \cite{Ablikim:2013wzq}, and $Z_c(3885)$ \cite{Ablikim:2013xfr}
were announced by BESIII by analyzing the $e^+e^-$ data at $\sqrt{s}=4.26$ GeV.
The $e^+e^-\to \gamma X(3872)$ process was explored in BESIII \cite{Ablikim:2013dyn}, where the $\sigma[e^+e^-\to \gamma X(3872)]\cdot B[X(3872)\to J/\psi\pi^+\pi^-]$ value was measured at $\sqrt{s}=4.009,\,4.229,\,4.260,\,4.360$ GeV \cite{Ablikim:2013dyn}.
These new experimental observations are important to further reveal the underlying structure of $Y(4260)$.

\subsection{$Y(4360)$ and $Y(4660)$}

By analyzing the $e^+e^- \to \pi^+ \pi^- \psi^\prime$ process via the Initial State Radiation, Belle observed two resonant structures $Y(4360)$ and $Y(4660)$ \cite{Wang:2007ea}, which were confirmed by BaBar \cite{Lees:2012pv}.

$Y(4360)$ was explained as a $3^3D_1$ charmonium or charmonium hybrid \cite{Li:2009zu,Ding:2007rg}, the radial excitation of $Y(4260)$ \cite{Qiao:2007ce},
a charmed baryonium \cite{Cotugno:2009ys}, the vector hybrid charmonium with strong coupling with $D^*\bar{D}_0$, $D_0\bar{D}^{*0}$ molecular state \cite{Kalashnikova:2008qr}, and a $2S$ $D^*\bar{D}_1$ molecular state \cite{Close:2010wq}. The situation of $Y(4360)$ is similar to that of $Y(4260)$. The above explanations must answer why $Y(4360)$ was not reported in the exclusive open-charm decay channels \cite{Abe:2006fj,Pakhlova:2008zza,Pakhlova:2007fq,Pakhlova:2009jv} and the $R$-value scan \cite{Burmester:1976mn,Brandelik:1978ei,Siegrist:1981zp,Osterheld:1986hw,Bai:1999pk,Bai:2001ct,CroninHennessy:2008yi,Ablikim:2009ad}. Thus, in Ref. \cite{Chen:2011kc}, the authors introduced the interference of $e^+e^-\to \psi(4160)/\psi(4415)\to \psi^\prime\pi^+\pi^-$ and the background contribution to
$e^+e^-\to \psi^\prime\pi^+\pi^-$, which is an important extension of Ref. \cite{Chen:2010nv}. They indicated that the $Y(4360)$ structure can be also reproduced well \cite{Chen:2011kc}.

The possible assignments to $Y(4660)$ are a $5^3S_1$ charmonium \cite{Ding:2007rg}, a baryonia with the flavor wave function $(|\Lambda_c^+\bar{\Lambda}_c^-\rangle+|\Sigma_c^0\bar{\Sigma}_c^0\rangle)/\sqrt{2}$ \cite{Qiao:2007ce},
a $f_0(980)\psi^\prime$ bound state \cite{Guo:2008zg}, and a P-wave tetraquark state \cite{Zhang:2010mw}.

\subsection{$Y(4630)$}

Belle announced the observation of an enhancement $Y(4630)$ near the $\Lambda_c\bar{\Lambda}_c$ threshold in
the $e^+e^-\to \Lambda_c\bar{\Lambda}_c$ process.

In Ref. \cite{vanBeveren:2008rt}, the enhancement structure near the $\Lambda_c\bar{\Lambda}_c$ threshold can be explained as the non-resonant signal, where the Resonance-Spectrum-Expansion (RSE) model was adopted. In addition, they also indicated that the Belle's data contains clear signals of $\psi(5S)$ and $\psi(4D)$ vector charmonia \cite{vanBeveren:2008rt}. By the screened potential, Li and Chao calculated the mass spectrum of charmonium, where $Y(4630)$ can correspond to $\psi(6S)$ state. The corresponding di-electron width is obtained, i.e. $\Gamma(\psi(6S)\to e^+e^-)=0.5$ keV. By this partial width, the width of $Y(4360)\to \Lambda_c^+\Lambda_c^-$ is extracted as $\Gamma(Y(4360)\to \Lambda_c^+\Lambda_c^-)=10$ MeV. A further study is needed to understand such a large baryonic decay width \cite{Li:2009zu}.
Cotugno {\it et al.} reanalyzed the data of $Y(4630)\to \Lambda_c\bar{\Lambda}_c$ and $Y(4660)\to \psi^\prime\pi^+\pi^-$, and indicated that these two observations can be due to the same state $Y_b$ with mass $m=4660.7\pm 8.7$ MeV and width $\Gamma=61\pm23$ MeV, where $Y_b$ is a charmed baryonium \cite{Cotugno:2009ys}. In Ref. \cite{Guo:2010tk}, the authors proposed that $Y(4630)$ and $Y(4660)$ are due to the same molecular state if taking into account the $\Lambda_c^+\Lambda_c^-$ final state interaction. In addition, the $\eta_c^\prime f_0(980)$ molecular state was predicted as the spin partner of $Y(4630)$ and $Y(4660)$ \cite{Guo:2010tk}.

\section{Two $X$ states from the double charm production}\label{sec4}

There exist two $X$ states from the double charm production, where $X(3940)$ \cite{Abe:2007jna,Abe:2007sya} and $X(4160)$ \cite{Abe:2007sya} can decay into charm meson pairs. The detailed information relevant to the production and decays of $X(3940)$ and $X(4160)$ includes
\begin{eqnarray}
e^+e^-\to \left\{\begin{array}{l}
X(3940)J/\psi\to \underline{D^*\bar{D}} J/\psi\\
X(4160)J/\psi\to \underline{D^{*+}D^{*-}} J/\psi
\end{array}\right.  .
\end{eqnarray}
In addition, the measured masses and widths of $X(3940)$ and $X(4160)$ are \cite{Abe:2007sya}
\begin{eqnarray*}
M_{X(3940)}&=&3942^{+7}_{-6}\pm6 \, \mathrm{MeV},\\
\Gamma_{X(3940)}&=&37^{+26}_{-15}\pm8 \, \mathrm{MeV},\\
M_{X(4160)}&=&4156^{+25}_{-20}\pm15 \, \mathrm{MeV},\\
\Gamma_{X(4160)}&=&139^{+111}_{-61}\pm21 \, \mathrm{MeV}.
\end{eqnarray*}
Since $X(3940)$ and $X(4350)$ are from the double charm production, thus their $C$ parities favor $C=+1$.

It is noted that there is no evidence that $X(3940)$ decays into $D\bar{D}$ \cite{Abe:2007sya}. Thus, we can exclude a scalar state assignment to $X(3940)$.
$X(3940)$ as a charmonium $\eta_c(3S)$ was proposed in Ref. \cite{Rosner:2005gf}. In the framework of the light cone formalism, Braguta {\it et al.} studied the $e^+e^-\to J/\psi X(3940)$ process assuming $X(3940)$ to be $\eta_c(3S)$ or one of the $2^3P_J$ states. Their results suggest that $X(3940)$ is a $\eta_c(3S)$ \cite{Braguta:2006py}. If explaining $X(3940)$ as $\eta_c(3S)$, there exists the low mass problem. The mass $X(3940)$ is lower than that predicted by the quenched potential model \cite{Barnes:2005pb} and the screened potential model \cite{Li:2009zu}. We also notice that the mass splitting between $X(3940)$ and $\psi(4040)$ is larger than that between $\eta_c^\prime$ and $\psi^{\prime\prime}$ \cite{Beringer:1900zz}. These unnatural properties still need to be understood. A different explanation, i.e., $X(3940)$ as a $2^1P_1$ charmonium, was proposed in Ref. \cite{Sreethawong:2013qua} by studying the decay behavior of $X(3940)$ as charmonium.
However, the $2^1P_1$ charmonium assignment to $X(3940)$ contradicts the estimate of the $C$ parity of $X(3940)$.

Since $X(4160)$ was observed in the $D^*\bar{D}^*$ channel but not in the $D\bar{D}$ and $D\bar{D}^*$ \cite{Abe:2007sya}, $X(4160)$ is a possible candidate of $\eta_c(4S)$ and $\chi_{c0}(3P)$ \cite{Chao:2007it}. If $X(4160)$ is $\eta_c(4S)$, $X(4160)$ cannot decay into $D\bar{D}$ while there exists $X(4160)\to D\bar{D}^*$. Thus, we need to explain why $X(4160)$ has the low mass and why $X(4160)\to D\bar{D}^*$ is suppressed. In Ref. \cite{Li:2009zu}, the masses of $\eta_c(4S)$ and $\chi_{c0}(3P)$ were predicted to be $4250$ MeV and $4131$ MeV, respectively, where $X(4160)$ favors $\chi_{c0}(3P)$. Under this assignment, the decay of $X(4160)$ into $D\bar{D}^*$ is forbidden while $X(4160)\to D\bar{D}$ is still allowed. Since experiment did not find $X(4160)\to D\bar{D}$, how to explain the suppression of $X(4160)\to D\bar{D}$ is crucial to test this possibility. A possible solution is that $X(4160)\to D\bar{D}$ is suppressed by the node effect \cite{Chao:2007it}.

Besides these conventional charmonium explanations, there are other discussions on the properties of $X(4160)$. In Ref. \cite{Molina:2009ct}, Molina and Oset proposed that $X(4160)$ can be a dynamically generated resonance from the vector-vector interaction, i.e., $X(4160)$ is a $D_s^*\bar{D}_s^*$ molecular state with $J^{PC}=2^{++}$.

More theoretical and experimental effort will be helpful to identify different explanations for $X(3940)$ and $X(4160)$.

\section{$X(3915)$, $X(4350)$ AND $Z(3930)$ produced by the $\gamma\gamma$ fusion}\label{sec5}

In the $\gamma\gamma$ fusion processes, experiments reported three charmoniumlike states, where $X(3915)$ \cite{Uehara:2005qd}, $X(4350)$ \cite{Shen:2009vs} and $Z(3930)$ \cite{Uehara:2009tx} decay into $D\bar{D}$, $J/\psi\phi$ and $J/\psi\phi$, respectively, which are summarized as
\begin{eqnarray}
\gamma\gamma &&\to \left\{\begin{array}{ll}
{X(3915)}\to \underline{D\bar{D}}\\
{X(4350)}\to \underline{J/\psi \phi} \\
{Z(3930)}\to \underline{J/\psi \omega}
\end{array}\right. .
\end{eqnarray}
Since the $\gamma\gamma$ fusion process is the filter of the $J^P=1^+$ state, thus the spin-parity quantum numbers of $X(3915)$, $X(4350)$ and $Z(3930)$ are either $0^+$ or $2^+$. In Table \ref{h2}, the measured masses and widths of $X(3915)$, $X(4350)$ and $Z(3930)$ are listed.
In Ref. \cite{Uehara:2009tx,Aubert:2010ab}, the angular distribution in the $\gamma\gamma$ center of mass frame shows $J^{PC}=2^{++}$ for $Z(3930)$. Thus, this fact indicates that $Z(3930)$ is a good candidate of the charmonium $\chi_{c2}^\prime$ \cite{Uehara:2009tx,Aubert:2010ab}.

\renewcommand{\arraystretch}{1.6}
\begin{table}[htb]
\caption{The resonance parameters of $X(3915)$, $X(4350)$ and $Z(3930)$. \label{h2}}
\begin{center}
\begin{tabular}{lll} \toprule[1pt]
State& Mass (MeV) & Width (MeV)\\\midrule[1pt]
$X(3915)$ \cite{Uehara:2005qd}& $3915\pm 3\pm2$  & $17\pm 10\pm3$\\
$X(4350)$ \cite{Shen:2009vs}& $4350^{+4.6}_{-5.1}\pm0.7$& $13.3^{+17.9}_{-9.1}\pm 4.1$\\
$Z(3930)$ \cite{Uehara:2009tx}& $3929\pm5\pm 2$&$29\pm10\pm2$\\
\bottomrule[1pt]
\end{tabular}
\end{center}
\end{table}

As shown in PDG \cite{Beringer:1900zz}, there are three P-wave states except for the radiative excitations, which are $\chi_{c0}(3415)$, $\chi_{c1}(3510)$ and $\chi_{c2}(3556)$. However, as for the first radial excitations of P-wave charmonia, $\chi_{c0}^\prime$ with $J^{PC}=0^{++}$ is missing, while $X(3872)$ and $Z(3930)$ can be regarded as $\chi_{c1}^\prime$ with $J^{PC}=1^{++}$ \cite{Meng:2005er,Suzuki:2005ha,Kalashnikova:2005ui,Danilkin:2010cc} and $\chi_{c2}^\prime$ with $J^{PC}=2^{++}$, respectively. Since the $\gamma\gamma$ fusion process is a good platform to create charmonium, it is natural to deduce whether the observed $X(3915)$ is a $\chi_{c0}^\prime$ state. In Ref. \cite{Liu:2009fe}, the authors studied this topic. The mass of $\chi_{c0}^\prime$ was predicted by the Godfrey-Isgur relativized potential model \cite{Barnes:2005pb}, which is close to the mass of $X(3915)$. In addition,
the coupling of $X(3915)$ and $D\bar{D}^*$ is fully forbidden if  $X(3915)$ is $\chi_{c0}^\prime$ while there exists the week interaction between $Z(3930)$ and $D\bar{D}^*$. However, the coupling between $X(3872)$ and $D\bar{D}^*$ via S-wave is very strong \cite{Kalashnikova:2005ui,Danilkin:2010cc}. These facts can answer why the mass difference between $X(3915)$ and $Z(3930)$ is smaller than that between
$X(3915)$ and $X(3872)$ \cite{Liu:2009fe}. Further study of two-body strong decay behavior of $X(3915)$ also supports the $\chi_{c0}^\prime$ assignment to $X(3915)$ \cite{Liu:2009fe}. After Ref. \cite{Liu:2009fe}, the BaBar Collaboration confirmed the
existence of the charmonium-like resonance $X(3915)$ decaying to
$J/\psi \omega$ with a spin-parity assignment $J^P=0^+$
\cite{Lees:2012xs}, i.e., they identified the signal being due to
the $\chi_{c0}(2P)$ which we have concluded in Ref. \cite{Liu:2009fe}.

As for $X(4350)$, the $\chi_{c2}^{\prime\prime}$ assignment was proposed in Ref. \cite{Liu:2009fe}. At first the mass of $X(4350)$ is consistent with the prediction by the Godfrey-Isgur relativized potential model \cite{Barnes:2005pb}. What is more important is that the calculated total width of $\chi_{c2}^{\prime\prime}$ can reproduce the width of $X(4350)$ \cite{Liu:2009fe}.

However, as the P-wave spin-triplet charmonium spectrum becomes complete, an urgent and crucial question
emerges out of the study of the first radial excitation of P-wave charmonia. The predicted mass of $\chi_{c0}(2P)$, as the first radial excitation of $\chi_{c0}(3415)$, is very close to that of $Z(3930)$ \cite{Kalashnikova:2005ui,Danilkin:2010cc} and above the $D\bar{D}$ threshold. Additionally, the $Z(3930)$ decay into $D\bar{D}$ occurs via the D-wave interaction, while the $\chi_{c0}(2P)\to D\bar{D}$ occurs via S-wave, where $\chi_{c0}(2P)\to D\bar{D}$ dominantly contributes to the total width of $\chi_{c0}(2P)$ \cite{Liu:2009fe}. Since $Z(3930)$ was already observed in the $D\bar{D}$ invariant mass spectrum of the $\gamma\gamma\to D\bar{D}$ process \cite{Uehara:2005qd,Aubert:2010ab}, we believe that $\chi_{c0}(2P)$ should exist in the data of the $D\bar{D}$ invariant mass spectrum since
we cannot find any suppression mechanism in the $\chi_{c0}(2P)$ production of $\gamma\gamma\to D\bar{D}$, where $\chi_{c0}(2P)$ and $Z(3930)$ have the same spatial wave function. However, the present experiment did not report any evidence of  $\chi_{c0}(2P)$ in the $\gamma\gamma\to D\bar{D}$ process \cite{Uehara:2005qd,Aubert:2010ab}, which obviously contradicts the above fact. It also becomes a new puzzle of studying P-wave higher charmonia. To solve this new puzzle, the authors in Refs. \cite{Chen:2012wy} proposed that the $Z(3930)$ structure may contain two P-wave higher charmonia ($\chi_{c0}^\prime$ and $\chi_{c2}^\prime$), which is supported by further analysis of the $D\bar{D}$ invariant mass spectrum and $\cos\theta$ distribution of $\gamma\gamma\to D\bar{D}$ \cite{Chen:2012wy}.


\section{Charged bottomoniulike and charmoniumlike states announced by Belle and BESIII}\label{sec6}

Belle observed two charged bottomoniumlike states $Z_b(10610)$ and $Z_b(10650)$
by studying the $e^+e^-$ annihilation near $\sqrt{s}=10.58$ GeV into hidden-bottom dipion channels \cite{Belle:2011aa}. In addition, their open-bottom decay modes were also reported by Belle \cite{Adachi:2012cx}. In 2013, the BESIII have made big progress in searching for the charged charmoniumlike states, which are $Z_c(3900)$ \cite{Ablikim:2013mio}, $Z_c(4025)$ \cite{Ablikim:2013emm}, $Z_c(4020)$ \cite{Ablikim:2013wzq} and $Z_c(3885)$ \cite{Ablikim:2013xfr}
from the analysis of $e^+e^-$ data at $\sqrt{s}=4.26$ GeV. The detailed decay information of these charged bottomoniumlike and charmoniumlike states is listed as follows
\begin{eqnarray*}
e^+e^-&&\to \left\{\begin{array}{l}
\left.
\begin{array}{l}
Z_b(10610) \pi^\mp\\
Z_{b}(10650)\pi^\mp\\
\end{array}\right\}\to \left\{\begin{array}{l}
\underline{\Upsilon(nS)\pi^\pm}\pi^\mp\,(n=1,2)\\
\underline{h_b(mP)\pi^\pm}\pi^\mp\,(m=1,2,3)\\
\underline{(B\bar{B}^*+c.c.)^\pm}\pi^\mp\\
\underline{(B^{*}\bar{B}^{*})^\pm} \pi^\mp \\
\end{array}\right.\\
\left.
\begin{array}{l}
Z_c(3900)\pi^\mp \to \underline{J/\psi\pi^\pm}\pi^\mp\\
Z_{c}(4025)\pi^\mp \to \underline{(D^{*} \bar{D}^{*})^{\pm}} \pi^\mp\\
Z_c(4020)\pi^\mp\to \underline{h_c\pi^\pm} \pi^\mp\\
Z_c(3885)\pi^+\to \underline{(D\bar{D}^*)}^-\pi^+ \\
\end{array}\right. 
\end{array}\right. .
\end{eqnarray*}

\subsection{$Z_b(10610)$ and $Z_b(10650)$}\label{sec6-1}

\begin{center}
\begin{figure}[htb]
\begin{tabular}{cccc}
\scalebox{0.64}{\includegraphics{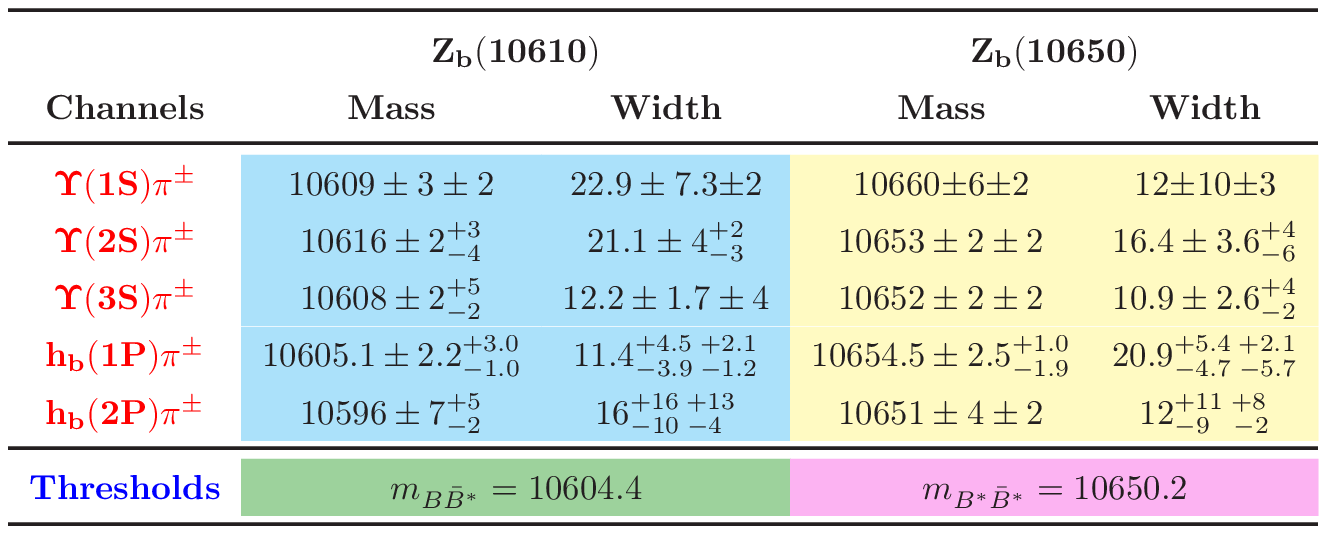}}
\end{tabular}
\caption{(color online). The masses and widths of $Z_b(10610)$ and $Z_b(10650)$ measured in the $\Upsilon(nS)\pi^\pm$ ($n=1,2,3$)
and $h_b(mP)\pi^\pm$ ($m=1,2$) invariant mass spectra \cite{Belle:2011aa}.
\label{zb} }
\end{figure}
\end{center}

In Fig. \ref{zb}, we collect the information of resonance parameters of $Z_b(10610)$ and $Z_b(10650)$ \cite{Belle:2011aa}. $Z_b(10610)$ and $Z_b(10650)$ have two typical peculiarities. Firstly, masses of $Z_b(10610)$ and $Z_b(10650)$ are close to the thresholds of $B\bar{B}^*$ and $B^*\bar{B}^*$, respectively. Secondly, $Z_b(10610)$ and $Z_b(10650)$ are charged states. Thus, $Z_b(10610)$ and $Z_b(10650)$ can be good candidates of exotic states.

Before the observations of $Z_b(10610)$ and $Z_b(10650)$, there have been many theoretical works which
focused on the molecular systems composed of $B^{(*)}$ and
$\bar{B}^{(*)}$ meson pair and indicated that there probably exist
loosely bound S-wave $B\bar{B}^{*}$ or $B^*\bar{B}^{*}$
molecular states \cite{Liu:2008fh,Liu:2008tn}.

As the first observed charged bottomoniumlike states,
$Z_b(10610)$ and $Z_b(10650)$ have attracted attention of many
theoretical groups. Bondar {\it et al.} discussed the special decay
behavior of the $J=1$ S-wave $B\bar{B}^*$ and $B^*\bar{B}^*$
molecular states under the heavy quark symmetry
\cite{Bondar:2011ev}. In Ref. \cite{Chen:2011zv}, the authors indicated
that the intermediate $Z_b(10610)$ and $Z_b(10650)$ contribution
to $\Upsilon(5S)\to \Upsilon(2S)\pi^+\pi^-$ naturally explains
Belle's previous observation of the anomalous
$\Upsilon(2S)\pi^+\pi^-$ production near the peak of
$\Upsilon(5S)$ at $\sqrt s=10.87$ GeV \cite{Abe:2007tk}, where the
resulting $d\Gamma(\Upsilon(5S)\to
\Upsilon(2S)\pi^+\pi^-)/dm_{\pi^+\pi^-}$ and
$d\Gamma(\Upsilon(5S)\to \Upsilon(2S)\pi^+\pi^-)/d\cos\theta$
distributions agree with Belle's measurement after inclusion of
these $Z_b$ states \cite{Chen:2011zv}.
Using a molecular bottomonium-like current in the
QCD sum rule calculation, Zhang {\it et al.}
\cite{Zhang:2011jja} tried to reproduce the masses of $Z_b(10610)$
and $Z_b(10650)$ .  In the chiral quark model,
Yang {\it et al.} calculated the mass
spectra of the S-wave $[\bar{b}q][b\bar{q}]$,
$[\bar{b}q]^*[b\bar{q}]$, $[\bar{b}q]^*[b\bar{q}]^*$, which shows that $Z_b(10610)$ and $Z_b(10650)$ are
good candidates of the S-wave $B\bar{B}^*$ and $B^*\bar{B}^*$
bound states \cite{Yang:2011rp}.
A non-exotic
explanation for $Z_b(10610)$ and $Z_b(10650)$ was proposed, where $Z_b(10610)$ and $Z_b(10650)$  are
interpreted as the orthogonal linear combinations of the $q\bar q$
and meson-meson states, namely $b\bar{b} + B\bar{B}^*$ and
$b\bar{b} + B^*\bar{B}^*$ \cite{Bugg:2011jr}, respectively.
Nieves
and Valderrama suggested the possible existence of two positive
C-parity isoscalar states: a $^3S_1-^3D_1$ state with a binding
energy of 90-100 MeV and a $^3P_0$ state located around 20-30 MeV
below the $B\bar{B}^*$ threshold \cite{Nieves:2011zz}. Unfortunately, the
quantum number of the above states does not match those of these
two charged $Z_b$ states. In addition,
Danilkin, Orlovsky and Simonov studied
the interaction between a light hadron and heavy quarkonium
via the transition to a pair of intermediate heavy mesons.
By the above coupled-channel effect, the authors discussed
the resonance structures close to the $B^{(*)}\bar B^\ast$
threshold \cite{Danilkin:2011sh}. Adopting the chromomagnetic interaction,
the authors of Ref. \cite{Guo:2011gu} discussed the possibility of
$Z_b(10610)$ and $Z_b(10650)$ being tetraquark states. In
contrast, the $b\bar b q\bar q$ tetraquark states were predicted
to be around $10.2 \sim 10.3$ GeV using the color-magnetic
interaction with the flavor symmetry breaking corrections
\cite{Cui:2006mp}, consistent with the values extracted from the QCD sum
rule approach \cite{Chen:2010ze}.

As specified in Ref. \cite{Chen:2011zv}, a future dynamical study
of the mass and decay mode of the S-wave $B\bar{B}^*$ and
$B^*\bar{B}^*$ molecular states are very desirable. Later, the authors of Refs. \cite{Sun:2011uh,Sun:2012zzd}
performed more thorough study on the $B\bar{B}^*$ and
$B^*\bar{B}^*$ systems using the One-Boson-Exchange (OBE) model. The numerical result shows
$Z_b(10610)$ and $Z_b(10650)$ can be explained as $B\bar{B}^*$ and
$B^*\bar{B}^*$ molecular states.

Assuming $Z_b(10610)$ and $Z_b(10650)$ to be the $B^{(*)}\bar{B}^{(*)}$ molecular states, Mehen and Powell obtained the line shapes in the vicinity of
$B^{(*)}\bar{B}^{(*)}$ thresholds and two-body decay rates of $Z_b(10610)$ and $Z_b(10650)$ and their partners under heavy quark symmetry \cite{Mehen:2011yh}. Later, the authors of \cite{Mehen:2013mva} calculated the differential distribution of $\Upsilon(5S)\to B^{(*)}\bar{B}^{(*)}\pi$, which can qualitatively describe the experimental data. They also found that the obtained angular distributions in the $\Upsilon(5S)\to Z_b(10610)/Z_b(10650)\pi$ decays are sensitive to the molecular character of $Z_b(10610)$ and $Z_b(10650)$ \cite{Mehen:2013mva}. In addition, $Z_b(10610)$ and $Z_b(10650)$ as the $B^{(*)}\bar{B}^{(*)}$ molecular state is also supported by investigating the
$h_b(nP)\pi^+$ invariant mass spectrum distributions of $\Upsilon(5S)\to h_b(nP)\pi^+\pi^-$ ($n=1,2$) \cite{Cleven:2011gp}. A further study of $Z_b(10610)$ and $Z_b(10650)$ was given in Ref. \cite{Cleven:2013sq}, where $Z_b(10610)$ and $Z_b(10650)$ decays into $\Upsilon(nS)\pi$, $h_b(mP)\pi$ and $\chi_{bJ}(mP)\gamma$ ($n=1,2,3$, $m=1,2$ and $J=0,1,2$) were calculated by the nonrelativistic effective field theory and under the assumption of the $B^{(*)}\bar{B}^{(*)}$ molecular state.

Besides the above theoretical interpretations, we also want to introduce a non-resonant explanation for $Z_b(10610)$ and $Z_b(10650)$. In Ref. \cite{Chen:2011pv}, the initial single pion emission (ISPE) mechanism was proposed to study the $\Upsilon(5S)$ hidden-bottom dipion decays. If the mass of higher bottomonium is larger than the sum of the masses of $B^{(*)}\bar{B}^{(*)}$ pair and pion, the corresponding bottomonium can have open-bottom decays associated with one pion production, where the emitted single pion plays an important role to make $B^{(*)}\bar{B}^{(*)}$ with low momenta. Hence, transformation of $B^{(*)}\bar{B}^{(*)}$ into the final state occurs via $B^{(*)}$ meson exchanges. By the ISPE mechanism, the
$Z_b(10610)$ and $Z_b(10650)$ structures can be naturally explained. In addition, this study also answers why Belle did not find the charged structure near the $B\bar{B}$ threshold in the $\Upsilon(nS)\pi^\pm$
and $h_b(mP)\pi^\pm$ channels \cite{Chen:2011pv}. Introduction of the ISPE mechanism in the $\Upsilon(5S)$ decay provides a unique perspective to understand the Belle's observation.
The ISPE mechanism was later applied to the hidden-charm dipion decays of higher charmonia and charmoniumlike states, which is due to the similarity between charmonium and bottomonium \cite{Chen:2011xk,Chen:2013bha}, where the charged charmoniumlike structures near the $D\bar{D}^{*}$ and $D^{*}\bar{D}^*$ thresholds were predicted. In the following subsection, we will further introduce these theoretical predictions combined with the experimental observations of $Z_c(3900)$, $Z_c(4025)$, $Z_c(4020)$ and $Z_c(3885)$.

\subsection{$Z_c(3900)$, $Z_c(4025)$, $Z_c(4020)$ and $Z_c(3885)$}

As the first charged charmoniumlike state announced by BESIII, $Z_c(3900)$ was observed in the $J/\psi\pi^\pm$ invariant mass spectrum of $e^+e^-\to J/\psi \pi^+\pi^-$ at $\sqrt{s}=4.26$ GeV \cite{Ablikim:2013mio}. $Z_c(3900)$ was later confirmed by Belle \cite{Liu:2013dau} in the same process and in Ref. \cite{Xiao:2013iha} in $e^+e^-\to J/\psi \pi^+\pi^-$ at $\sqrt{s}=4.16$ GeV.
The mass and width of $Z_c(3900)$ from different experiments are listed in Table \ref{z3900}. The typical property of $Z_c(3900)$ is that it is near the $D\bar{D}^*$ threshold, which is the reason why it can be a good candidate of an exotic state.

\renewcommand{\arraystretch}{1.6}
\begin{table}[htb]
\caption{The mass and width of $Z_c(3900)$ measured by different experiments. \label{z3900}}
\begin{center}
\begin{tabular}{ccc} \toprule[1pt]
Experiments & Mass (MeV) & Width (MeV)\\\midrule[1pt]
BESIII \cite{Ablikim:2013mio}& $3899.0\pm 3.6\pm 4.9$  & $46\pm 10\pm 20$\\
Belle \cite{Liu:2013dau}& $3894.5\pm6.6\pm4.5$& $63\pm24\pm26$\\
Xiao {\it et al.} \cite{Xiao:2013iha}& $3886\pm4\pm 2$&$37\pm4\pm8$\\
\bottomrule[1pt]
\end{tabular}
\end{center}
\end{table}

In Sec. \ref{sec6-1}, we mentioned the prediction by the ISPE mechanism. In Ref. \cite{Chen:2011xk}, the decays of $Y(4260)$ into $J/\psi \pi^+\pi^-$, $\psi^\prime \pi^+\pi^-$ and $h_c(1P)\pi^+\pi^-$ were studied by the ISPE. The authors explicitly indicated that there exist charged charmoniumlike structures near $D\bar{D}^*$ and $D^*\bar{D}^*$ thresholds in the corresponding $J/\psi\pi^\pm$, $\psi^\prime\pi^\pm$ and $h_c(1P)\pi^\pm$ invariant mass spectra. The observation of $Z_c(3900)$ confirmed the above theoretical prediction.

After the discovery of $Z_c(3900)$, the authors of Ref. \cite{Chen:2013coa} studied the the distributions of $J/\psi\pi^\pm$ and $\pi^+\pi^-$ invariant  mass spectra of $Y(4260)\to J/\psi\pi^+\pi^-$ by taking into account the interference effects of the ISPE mechanism with two other decay modes. The numerical result shows that the $Z_c(3900)$ structure can be well reproduced \cite{Chen:2013coa}.

As emphasized above, the peculiarity of $Z_c(3900)$ makes it be a good exotic state candidate.
Before the observation of $Z_c(3900)$, the authors of Refs. \cite{Sun:2011uh,Sun:2012zzd} predicted the existence of the $D\bar{D}^*$ and $D^*\bar{D}^*$ molecular states by the OBE model.

$Z_c(3900)$ has also stimulated further study of whether it is an exotic state.
The authors of Ref. \cite{Wang:2013cya} suggested that $Z_c(3900)$ is a charged $D\bar{D}^*+\bar{D}D^*$ molecular state with $J^P=1^+$, while $Y(4260)$ is a $\bar{D}D_1(2420) +D\bar{D}_1(2420)$ molecular state. Using the heavy quark symmetry and assuming $X(3872)$ and $Z_b(10610)$ as $D\bar{D}^*$ and $B\bar{B}^*$ molecular states, respectively, Guo {\it et al.} obtained a series of hadronic molecules composed of heavy mesons. One of their results can correspond to $Z_c(3900)$, which indicates the possibility of  $Z_c(3900)$ as an isovector $D\bar{D}^*$ molecular state \cite{Guo:2013sya}.
In Ref. \cite{Maiani:2004vq}, Maiani {\it et al.} predicted a charged tetraquark state with mass around 3882 MeV when proposing the tetraquark explanation for $X(3872)$. By fitting the BESIII \cite{Ablikim:2013mio} and Belle \cite{Liu:2013dau} data, the authors of Ref. \cite{Faccini:2013lda} pointed out that $Z_c(3940)$ can correspond to the above tetraquark state. Since another tetraquark state with mass 3755 MeV was also predicted in Ref. \cite{Maiani:2004vq}, the discussion on the possibility of having it in the BESIII and Belle data was also given in Ref. \cite{Faccini:2013lda}.
Voloshin \cite{Voloshin:2013dpa} discussed the possibilities of $Z_c(3900)$ as the $D\bar{D}^*$ molecular state, hadro-charmonium and tetraquark state.
Several QCD sum rule calculations relevant to $Z_c(3900)$ were performed in Refs. \cite{Cui:2013yva,Zhang:2013aoa,Dias:2013xfa,Wang:2013vex}.
Cui {\it et al.} obtained mass $(3.91\pm0.19)$ GeV by a $D^*\bar{D}$ molecular state current with $I^GJ^P=1^+1^+$, where they consider the contribution up to dimension six in the operator product expansion at the leading order in $\alpha_s$ \cite{Cui:2013yva}. Later, Zhang carried out an improved QCD
sum rule study of $Z_c(3900)$ \cite{Zhang:2013aoa} and claimed that their result supports the S-wave $D\bar{D}^*$ molecular state assignment. Using the three-point QCD sum rule and identifying $Z_c(3900)$ as the tetraquark partner of $X(3872)$, Dias {\it et al.} calculated the coupling constants of $Z_c^+(3900)J/\psi\pi^+$, $Z_c^+(3900)\eta_c\rho^+$ and $Z_c^+(3900)D^+\bar{D}^{*0}$ interactions. Further they obtained the total width of $Z_c(3900)$, which is consistent with the experimental data \cite{Dias:2013xfa}. Very recently, Wang and Huang indicated that $Z_c(3900)$ can be a $1^{+-}$
diquark-antidiquark type tetraquark state by the analysis with the QCD sum rule \cite{Wang:2013vex}.
Besides suggesting $Y(4260)$ as the lowest $1^{--}$ charmonium hybrid, Braaten claimed that the observed $Z_c(3900)$ is a tetraquark state, i.e., a $0^{-+}$ $c\bar{c}$ pair plus an isovector $q\bar{q}$ pair.
There are some theoretical studies on the $Z_c(3900)$ decays \cite{Braaten:2013boa}. By the effective Lagrangian approach, the authors of Ref. \cite{Dong:2013iqa} predicted the hidden-charm decay widths of $Z_c(3900)\to \psi(nS)\pi, h_c(mP)\pi$ under the $D\bar{D}^*$ molecular state assumption. In Ref. \cite{Ke:2013gia}, Ke {\it et al.} calculated the partial decay widths of $Z_c(3900)$ as a $D\bar{D}^*$ molecular state into $J/\psi\pi$, $\psi^\prime \pi$ and $\eta_c\rho$ by the light front model and they found that $Z_c(3900)\to D\bar{D}^*$ is rather small and $\Gamma(Z_c(3900)\to \psi^\prime \pi)>\Gamma(Z_c(3900)\to J/\psi \pi)$.
A Lattice study of $Z_c(3900)$ was performed in Ref. \cite{Prelovsek:2013xba} by adopting the meson-meson type interpolators, where they did not find a candidate for $Z_c(3900)$ with $I(J^{PC})=1(1^{+-})$.
Recently, Lin, Liu and Xu further explored the possibility to discover $Z_c(3900)$ via the meson photoproduction process assuming $Z_c(3900)$ as the $D\bar{D}^*$ molecular state \cite{Lin:2013mka}.

As a new charged charmoniumlike state near the $D^*\bar{D}^*$ threshold, $Z_c(4025)$ was observed in the $e^+e^-\to (D^*\bar{D}^*)^\pm \pi^\mp$ process at $\sqrt{s}=4.26$ GeV.
The mass and width of $Z_c(4025)$ are $4026\pm2.6\pm3.7$ MeV and $24.8\pm5.6\pm7.7$ MeV, respectively \cite{Ablikim:2013emm}. Before the observation of $Z_c(4025)$, there were some theoretical predictions of
charged charmoniumlike state around the $D^*\bar{D}^*$ threshold. In Ref. \cite{Sun:2011uh,Sun:2012zzd},
an isovector $D^*\bar{D}^*$ molecular state was predicted by using the OBE model. By the ISPE mechanism,  Chen and Liu indicated that there exist charged charmoniumlike structures near the $D^*\bar{D}^*$ threshold in the $J/\psi\pi^\pm$, $\psi^\prime\pi^\pm$ and $h_c(1P)\pi^\pm$ invariant mass spectra of $Y(4260)\to J/\psi \pi^+\pi^-, \psi^\prime \pi^+\pi^-,h_c(1P)\pi^+\pi^-$ \cite{Chen:2011xk}.
Chen, Liu and Matsuki later applied the ISPE mechanism to study $e^+e^-\to (D^{(*)}\bar{D}^{(*)})^\pm \pi^\mp$ processes, where the charged charmoniumlike structures near the $D\bar{D}^*$ and $D^*\bar{D}^*$ thresholds appear in the corresponding $D^{(*)}\bar{D}^{(*)}$ invariant mass spectrum, one of which can correspond to $Z_c(4025)$ \cite{Chen:2012yr}.

Similar to the situation of $Z_c(3900)$, the observation of $Z_c(4025)$ has also inspired the discussions of the underlying mechanism behind this novel phenomenon. In Ref. \cite{He:2013nwa}, the authors studied the loosely bound $D^*\bar{D}^*$ system, and pointed out that $Z_c(4025)$ can be an ideal $D^*\bar{D}^*$ molecular state with $I^G(J^P)=1^+(1^+)$. This quantum number assignment is due to the assumption that  $Z_c(4025)$ and $Z_c(4020)$ \cite{Ablikim:2013wzq} are the same state \cite{He:2013nwa}, where $Z_c(4020)$ was reported in the $h_c\pi^\pm$ invariant mass spectrum of $e^+e^-\to h_c\pi^+\pi^-$ at $\sqrt{s}=4.26$ GeV \cite{Ablikim:2013wzq}.
The mass and width of $Z_c(4020)$ are $4022.9\pm 0.8\pm 2.7$ MeV and $7.9\pm 2.7\pm 2.6$ MeV \cite{Ablikim:2013wzq}.
Further the decay behaviors of these $D^*\bar{D}^*$ molecular states with $0^+(0^{++})$, $0^+(2^{++})$, and $0^-(1^{+-})$ were predicted in the heavy quark limit \cite{He:2013nwa}. By the approach of the QCD sum rule, Cui {\it et al.} suggested that $Z_c(4025)$ can be a $D^*\bar{D}^*$ molecular state with $J^P=1^+$ \cite{Cui:2013vfa}. The same conclusion was also obtained in Refs.  \cite{Chen:2013omd}. In Ref. \cite{Qiao:2013dda}, Qiao and Tang calculated the masses by the tetraquark $[cu][\bar{c}\bar{d}]$ currents with $J^P=1^-$ and $2^+$. They suggested that $Z_c(4025)$ is a $J^P=2^+$ tetraquark state. Using the QCD sum rule, Khemchandani {\it et al.} obtained the masses of $1^+$ and $2^+$ states with the $\bar{D}^{*0}D^{*+}$ molecule currents, both of which are consistent with the experimental data of $Z_c(4025)$ \cite{Khemchandani:2013iwa}. The above QCD sum rule studies of $Z_c(4025)$ give different results on the $Z_c(4025)$ structure.

Besides studying $Z_c(4025)$ due to exotic state explanations, there exist other proposals to $Z_c(4025)$, i.e., the non-resonant explanation for $Z_c(4025)$. Being combined with the experimental data of the $\pi^-$ recoil mass spectrum \cite{Ablikim:2013emm}, Wang {\it et al.} investigated $Y(4260)\to (D^*\bar{D}^*)^- \pi^+$ decay via the ISPE mechanism and found that the $Z_c(4025)$ structure can be reconstructed \cite{Wang:2013qwa}. Later, the authors in Ref. \cite{Torres:2013lka} analyzed the data of $e^+e^-\to (D^*\bar{D}^*)^\pm \pi^\mp$. They indicated that the BESIII data can be interpreted without introducing $Z_c(4025)$ resonance.

According to the above review of the theoretical status of $Z_c(4025)$, we also notice an interesting fact. The conclusion of whether $Z_c(4025)$ and $Z_c(4020)$ are the same state is crucial since it gives different constraints on the quantum numbers of $Z_c(4025)$ and $Z_c(4020)$. If only making a comparison between $Z_c(4025)$ \cite{Ablikim:2013emm} and $Z_c(4020)$ \cite{Ablikim:2013wzq} on the measured masses and widths,
the width of $Z_c(4025)$ is different from that of $Z_c(4020)$. However, it is not enough to conclude whether $Z_c(4025)$ and $Z_c(4020)$ are the same or not only by the width difference between $Z_c(4025)$ and $Z_c(4020)$. Further experimental information like the measurement of the angular distribution will clarify this puzzle.

\section{Summary}\label{sec7}

Over the past decade, the family of charmoniumlike and bottomoniumlike states has become more and more abundant due to the experimental development. It is a research topic full of opportunities and challenges for theorists as well as experimentalists to reveal the inner mechanisms originating from these novel and complicated phenomena. With the experimental progress, theorists have paid more attention to these observations by proposing different explanations. In this review article, we briefly summarize the progress and recent developments on theoretical study of $XYZ$ new particles combined with the experimental status.

By giving this review, we also learn some valuable lessons and revelations:

\begin{itemize}
\item There exist different theoretical interpretations to each and every experimental observation. Thus, how to further distinguish them is very crucial, which requires the joint efforts of theorists and experimentalists.

\item Although the observed $XYZ$ new particles have stimulated extensive study of whether
these new particles are all exotic states, we cannot exactly identify some observed charoniumlike states with exotic ones. Before giving definite conclusion of identifying an exotic state, we should exhaust all possibilities in the conventional mechanisms to explain these experimental observations.

\item The conclusion depends on the model. For example, different potential models give different mass spectra of charmonium family. Sometimes the predicted properties for a concrete exotic state by various models are different from each other. This fact shows that the phenomenological models reflect only a part of true physics picture. Hence, these new experimental observations can provide a good platform to further develop the phenomenological models.

\end{itemize}

Up to now, these reported $XYZ$ new particles have opened a new field of particle physics and almost covered all particle physics experiments within the range in between 2 to 10 GeV, which include Belle, BaBar, CDF, D$\varnothing$, LHCb, CMS, and BESIII. With the run of the forthcoming experiments (BelleII and PANDA), we can expect that there will be more experimental discoveries of $XYZ$ states. In the next decade, it will be an exciting and challenging time both for experimentalists and theorists.

\vfil
\section*{Acknowledgements}
The author thanks Professor Tao Huang for suggesting the topic and all my collaborators for the enjoyable
collaborations. I also would like to thank Professor Takayuki Matsuki for reading the manuscript and suggestive comments.
Because of his personal research interest, the author apologizes to those whose papers and contributions were not mentioned in this work.
This project is supported by the National Natural Science
Foundation of China under Grants No. 11222547, No. 11175073 and No. 11035006, the Ministry of Education of China
(FANEDD under Grant No. 200924, SRFDP under Grant No.
2012021111000, and NCET), the Fok Ying Tung Education Foundation
(No. 131006).

\end{document}